\documentclass[prd,aps,superscriptaddress]{revtex4}
\usepackage{graphicx}% Include figure files
\usepackage{dcolumn}% Align table columns on decimal point
\usepackage{bm}% bold math

\begin{document}
\voffset = 0.3 true in
\topmargin = -1 true in % for Mac tetex

\title{Short Range Structure in the $X(3872)$}

\author{Eric S. Swanson}
\affiliation{
Department of Physics and Astronomy, University of Pittsburgh,
Pittsburgh PA 15260}
\affiliation{
Jefferson Lab, 12000 Jefferson Ave,
Newport News, VA 23606}

\vskip .5 true cm
\begin{abstract}
It is proposed that the newly discovered $X(3872)$ is a $J^{PC} = 1^{++}$
$D^0\bar D^{0*}$ hadronic resonance stabilized by admixtures of $\omega J/\psi$ and
$\rho J/\psi$. A specific model of the state is constructed and tests
of its internal structure are suggested via the predicted decay modes $D^0\bar D^0\pi^0$, 
$D^0\bar D^0\gamma$, $\pi^+\pi^- J/\psi$, and $\pi^+\pi^-\pi^0 J/\psi$.
\end{abstract}

%\pacs{}
\maketitle

\section{Introduction}

The Belle collaboration has recently announced\cite{B} the discovery of a 
resonance, $X(3872)$, in the $\pi^+\pi^-J/\psi$ subsystem of the process 

\begin{equation}
B^\pm \to K^\pm \pi^+\pi^- J/\psi
\end{equation}
at a mass of $3872.0 \pm 0.6 \pm 0.5$ MeV and with a width

\begin{equation}
\Gamma < 2.3 {\rm MeV} \ \ (95\% {\rm C.L.}).
\end{equation}
This state, which has been
confirmed by the CDF collaboration\cite{CDF}, has attracted some attention
because of its unusual properties. Specifically, the state appears to be too heavy to be
a 1D charmonium state and too light to be 2P charmonium or a $c\bar c$ hybrid.
See Ref. \cite{BG} for a detailed assessment of possible charmonium assignments and decay modes.

Alternatively, the proximity of the state to $D \bar D^*$ threshold
strongly suggests that the $X$ may be a weakly bound $D\bar D^*$ 
resonance\cite{NAT2,CP,BK,CYW,PS}, sometimes
called a mesonic `molecule'  or a `deuson'\cite{general}. This is an old idea which 
has been
applied to a variety of mesons with unusual characteristics such as the 
$\psi(4040)$\cite{psi}, $f_1(1420)$\cite{E,ess}, $\eta(1440)$\cite{NAT}, 
$f_J(1720)$\cite{DSB}, $a_0(980)$, and $f_0(975)$\cite{scalars,ess}.

In this note I assume that the $X(3872)$ is indeed a $D\bar D^*$ resonance and 
present a detailed analysis of its expected properties based on a simple model
of quark interactions. This model incorporates the nonrelativistic quark model with
additional dynamics due to pion exchange. The idea is to capture the predictive
power of a microscopic formalism of short range quark dynamics along   with  the important long
range dynamics mediated by pion exchange processes.
Versions of this idea  have been applied
to baryon-baryon interactions since the 1980's\cite{RGM}, where, of course, 
pion exchange is
of fundamental importance; another variant
has recently enjoyed some vogue in baryon physics\cite{GR}.

It is natural to expect that the putative $D\bar D^*$ bound state is in a relative
S-wave since this is typically where inter-hadron forces are strongest. 
In this case pion-mediated interactions (see below) favour 
the isoscalar channel, which in turn implies a $J^{PC}= 1^{++}$ state. I will 
therefore henceforth refer to the bound state interpretation of the $X(3872)$ 
as the $\hat\chi_{c1}(3872)$.
The remainder of this paper focusses on the properties of this state. 

Although pion exchange forces dominate the structure of the $\hat\chi_{c1}$ (in analogy to
the deuteron), short range quark dynamics are present and assist in binding the
$\hat \chi_{c1}$ via mixing to hidden charm vector-$J/\psi$ states. Indeed, $\omega J/\psi$ and
$\rho J/\psi$ are very nearly degenerate with $D\bar D^*$ and one must expect 
some admixture of these states -- an effect which will be strongly enhanced by the
near-zero energy denominator.  Such mixing is also important in driving possible
decay modes of the $\hat\chi_{c1}$ and is therefore central to determining its properties.
Finally, the binding energy of the $\hat\chi_{c1}$ is comparable to mass differences in
the available charge channels and one can expect strong isospin violating effects
in this resonance.
This heretofore unexplored dynamics is thoroughly examined in the following.
Detailed predictions of binding energies and branching fractions are 
presented along with possible experimental tests of $\hat\chi_{c1}$ structure.

\section{$D\bar D^*$ Dynamics}

Long range pion exchange effects are expected to dominate the physics of a 
weakly bound state
such as the $\hat\chi_{c1}$. Nevertheless, as discussed above, short range quark interactions
can give rise to important mixing effects.  We therefore consider a model which
appends pion exchange dynamics to the nonrelativistic quark model. The model is used
to extract effective interactions for $D\bar D^*-D\bar D^*$, $D\bar D^*-\omega J/\psi$, and 
$D\bar D^*-\omega J/\psi$ scattering. These interactions are then employed in a 
nonrelativistic coupled channel Schr\"odinger equation to extract bound
state properties (one expects the nonrelativistic
formalism to be accurate for weakly bound states of relatively massive components
as is the case with the $\hat\chi_{c1}$).

\subsection{Quark Exchange Induced Effective Interaction}

The quark model employed here assumes nonrelativistic quark dynamics mediated
by an instantaneous confining interaction and a short range spin-dependent interaction
motivated by one gluon exchange. The colour structure is taken to be the 
quadratic form of  perturbation theory. This is an important assumption for 
multiquark dynamics which has received support from recent lattice 
computations for both confinement\cite{Bali} and multiquark interactions\cite{BB}.
The final form of the interaction is thus taken to be

\begin{equation}
\label{Vij}
\sum_{i<j}{\bm{\lambda}(i) \over 2}\cdot {\bm{\lambda}(j) \over 2} \left \{
{\alpha_s \over r_{ij}} - {3\over 4} br_{ij}
- {8 \pi \alpha_s \over
3 m_i m_j } \bm{S}_i \cdot \bm{S} _j \left ( {\sigma^3 \over
\pi^{3/2} } \right ) e^{-\sigma^2 r_{ij}^2}
\right \},
\end{equation}
where ${\bm{\lambda}}$ is a colour Gell-Mann matrix, $\alpha_s$ is
the strong coupling constant, $b$ is the string tension, $m_i$ and
$m_j$ are the interacting quark or antiquark masses, and $\sigma$ is a
range parameter in a regulated  spin-spin hyperfine
interaction.  The parameters used were $\alpha_s = 0.59$, $b = 0.162$ GeV$^2$,
$\sigma = 0.9$ GeV, and $0.335$, $0.55$, and $1.6$ GeV for up, strange, and
charm quark masses respectively. Relevant meson masses obtained from this model are
$\rho = 0.773$ GeV, 
$J/\psi = 3.076$ GeV, $D = 1.869$ GeV, and $D^* = 2.018$ GeV, in good agreement
with experiment.

Meson-meson interactions are obtained by computing the Born order scattering amplitude
for a given process\cite{ess,BS}. Because of the colour factors in Eq. \ref{Vij} this amplitude
necessarily involves an exchange of quarks between the interacting mesons. Thus the 
leading order $D\bar D^*$ interaction  couples $D\bar D^*$ with hidden charm states
such as $\rho J/\psi$ and $\omega J/\psi$. This amplitude may be unitarized by 
extracting an effective potential and iterating it in a Schr\"odinger equation\cite{ess}. 
The method has been successfully applied to a variety of processes such 
as $KN$ scattering\cite{KN}
and  $J/\psi$ reactions relevant to RHIC physics\cite{WSB}. It has even 
proven surprisingly useful for relativistic (and chiral) reactions such as 
$\pi\pi$ scattering\cite{BS,ess}.

%\begin{figure}[h]
%\includegraphics[angle=0,width=10cm]{qex1.ps}
%\caption{\label{qex} Quark Exchange Amplitudes}
%\end{figure}

\begin{figure}[h]
\includegraphics[angle=270,width=8cm]{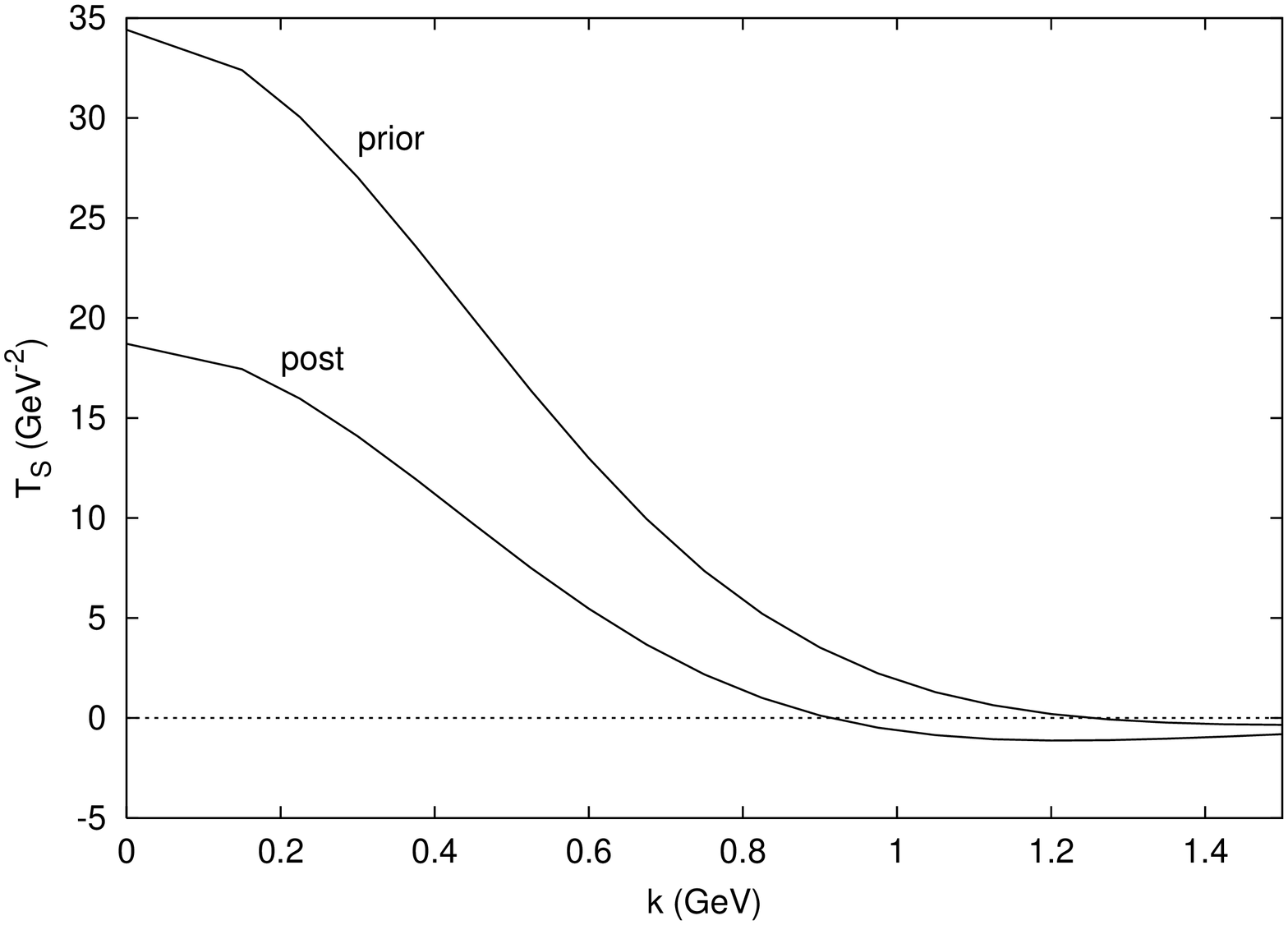}
\includegraphics[angle=270,width=8cm]{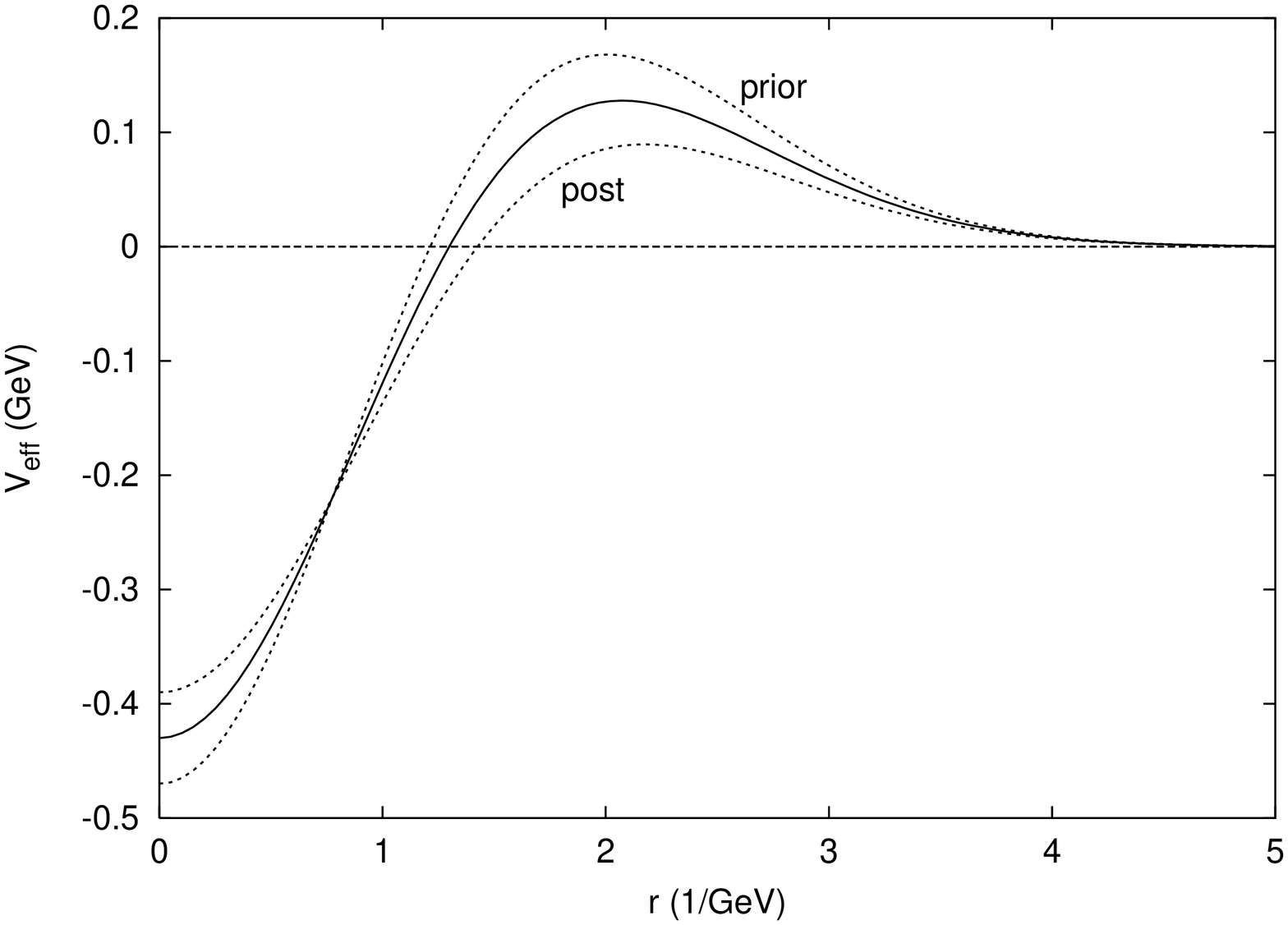}
\caption{\label{TC} (left) S-Wave Scattering Amplitude for $D\bar D^* \to \omega J/\psi$.
(right) Effective Potential for $D\bar D^* \to \omega J/\psi$ }
\end{figure}

The S-wave Born order scattering amplitude for $D\bar D^* - \omega J/\psi$ scattering
is shown in Fig. \ref{TC}. Here $D \bar D^*$ refers to the isoscalar positive charge parity state $1/\sqrt{2}(D\bar D^* + \bar D D^*)^0_S$. The scattering amplitude is
dominated by the confinement interaction of Eq. \ref{Vij} (this is in contrast to 
light meson scattering which is dominated by the hyperfine interaction). An effective
potential is extracted by equating the scattering amplitude to that obtained for
point-like mesons interacting via an arbitrary S-wave potential. It is convenient
to parameterize this potential as a sum of gaussians:

\begin{equation}
V_q = \sum_i a_i {\rm e}^{-r^2/2b_i^2}.
\label{Vq}
\end{equation}

The fit to the quark level
amplitude is illustrated in Fig. \ref{TC} (left panel) and the resulting potential
is shown in Fig. \ref{TC} (right panel). The distinctive 
``mermaid potential''
seen here is due to destructive interference between diagrams in the quark level 
amplitude. Thus details of the potential are sensitive to the assumed microscopic
interaction, however, its general shape and strength are quite robust\cite{ess}.

The derived parameters of Eq. \ref{Vq} were  $a_1 = 6.35$ GeV, $b_1 = 1.166$ GeV$^{-1}$ and
$a_2 = -6.82$ GeV, $b_2 = 1.096$ GeV$^{-1}$. These parameters were obtained for
the confinement portion of the ``prior'' form of the scattering amplitude. The ``post''
form yields $a_1 = 3.82$ GeV, $b_1 = 1.20$ GeV$^{-1}$, $a_2 = -4.21$ GeV, and $b_2 = 1.125$ GeV$^{-1}$. Post and prior forms of a scattering amplitude refer to different schemes for
constructing the time evolution operator which exist in the scattering of composite
systems. In principle these
give rise to the same scattering amplitude, but approximations and inaccurate
wavefunctions can cause slight differences as indicated in the figure.
We employ the average potential indicated by the solid line of Fig. \ref{TC} (right) 
in the following.
Finally, the isovector $D\bar D^*$- $\rho J/\psi$ effective potential is identical to its
isoscalar analogue.

The mermaid form of the effective potential implies that quark exchange effects
can cause binding in the coupled $D\bar D^*$, $\omega J/\psi$ or $\rho J/\psi$ 
systems; however, direct computations indicate that the potential depth is not 
sufficient to form a resonance. We therefore turn to an examination of pion
exchange induced dynamics in the $D\bar D^*$ system.

\subsection{Pion Exchange Induced Effective Interaction}

I choose to follow the method of T\"ornqvist\cite{NAT} in constructing an effective
pion-induced interaction. This is based  on a microscopic quark-pion interaction
familiar from nuclear physics:

\begin{equation}
{\cal L} = {g\over \sqrt{2} f_\pi} \int d^4x \bar\psi(x) \gamma^\mu \gamma_5 \tau^a \psi(x) \partial_\mu \pi^a(x).
\label{piq}
\end{equation}
Here $f_\pi = 92$ MeV is the pion decay constant, $\tau$ is an SU(2) flavour generator,
and $g$ is a coupling to be determined. The effective potential is derived by projecting
the quark level interactions onto hadronic states in the nonrelativistic limit. In 
the case of pseudoscalar-vector states one obtains\cite{NAT}

%\begin{equation}
%V_{\pi qq} = -{g \over \sqrt{2} f_\pi} \int d^3x (\sigma\cdot\nabla) (\tau\cdot \pi)
%\end{equation}

\begin{equation}
V_{\pi} = - \gamma V_0 \left[ \pmatrix{1 & 0 \cr 0 & 1}C(r) + \pmatrix{0 & -\sqrt{2} \cr 
-\sqrt{2} & 1 } T(r) \right]
\label{Vpi}
\end{equation}
where

\begin{equation}
C(r) = {\mu^2\over m_\pi^2} {{\rm e}^{-\mu r} \over m_\pi r},
\label{C}
\end{equation}
\begin{equation}
T(r) = C(r)\left( 1 + {3\over \mu r} + {3\over (\mu r)^2} \right),
\label{T}
\end{equation}
and
\begin{equation}
V_0 \equiv {m_\pi^3\over 24 \pi}{g^2 \over f_\pi^2} \approx 1.3 {\rm MeV}.
\end{equation}
The matrix elements refer to S- and D-wave components of the pseudoscalar-vector
state in analogy with the deuteron.
The strength of the interaction has been fixed by comparing to the $\pi NN$ coupling 
constant via the 
relationship $g_{\pi NN}^2/4\pi = 25/18\cdot m_\pi^2 g^2/f_\pi^2$. This allows a
prediction of the $D^*$ decay width which is in good agreement with experiment\cite{NAT}.
The parameter $\mu$ is typically the pion mass, however, one can incorporate recoil
effects in the potential by setting $\mu^2 = m_\pi^2 - (m_V-m_{pS})^2$. The results
presented here are insensitive to the value of $\mu$ and I take $\mu = 130$ MeV in the
following.
Finally, the coupling $\gamma$ is a spin-flavour matrix element which takes on the 
following values: $\gamma = 3$ for $I=0$, $C=+$; $\gamma = 1$ for $I=1$, $C=-$;
$\gamma = -1$ for $I=1$, $C=+$; and $\gamma = -3$ for $I=0$, $C=-$. Thus the isoscalar
positive charge parity channel is the most likely to form bound states and subsequent
discussion focusses on it.

The potential of Eq. \ref{T} is an illegal quantum mechanical operator and
must be regulated, typically with a dipole form factor. The regulator scale, $\Lambda$ 
may be fixed by comparison with nuclear physics; for example $NN$ interactions
yield preferred values for $\Lambda$ in the range 0.8 GeV to 1.5 GeV depending on model
details. Alternatively, 
reproducing the deuteron binding energy requires $\Lambda \approx 0.8$ GeV. 
T\"ornqvist has employed an intermediate value of $\Lambda = 1.2$ GeV which is 
appropriate for $D$ mesons and this is taken as the 
canonical cutoff in the following.

Integrating the coupled S/D wave system for the $1^{++}$ $B\bar B^*$ system yields a 
bound state of mass 10562 MeV, in agreement with Ref. \cite{NAT}. Similarly a $0^{-+}$
$B\bar B^*$ bound state of mass 10545 MeV arises from this formalism.
Unfortunately,
$D$ mesons are sufficiently light that the $D\bar D^*$ system does not
bind with canonical parameters. However, the combined pion and quark
induced effective interactions are sufficient to cause binding. The properties
of this bound state are explored in the next section.

\section{Properties of the $\hat \chi_{c1}(3872)$}

\begin{figure}[h]
\includegraphics[angle=270,width=8cm]{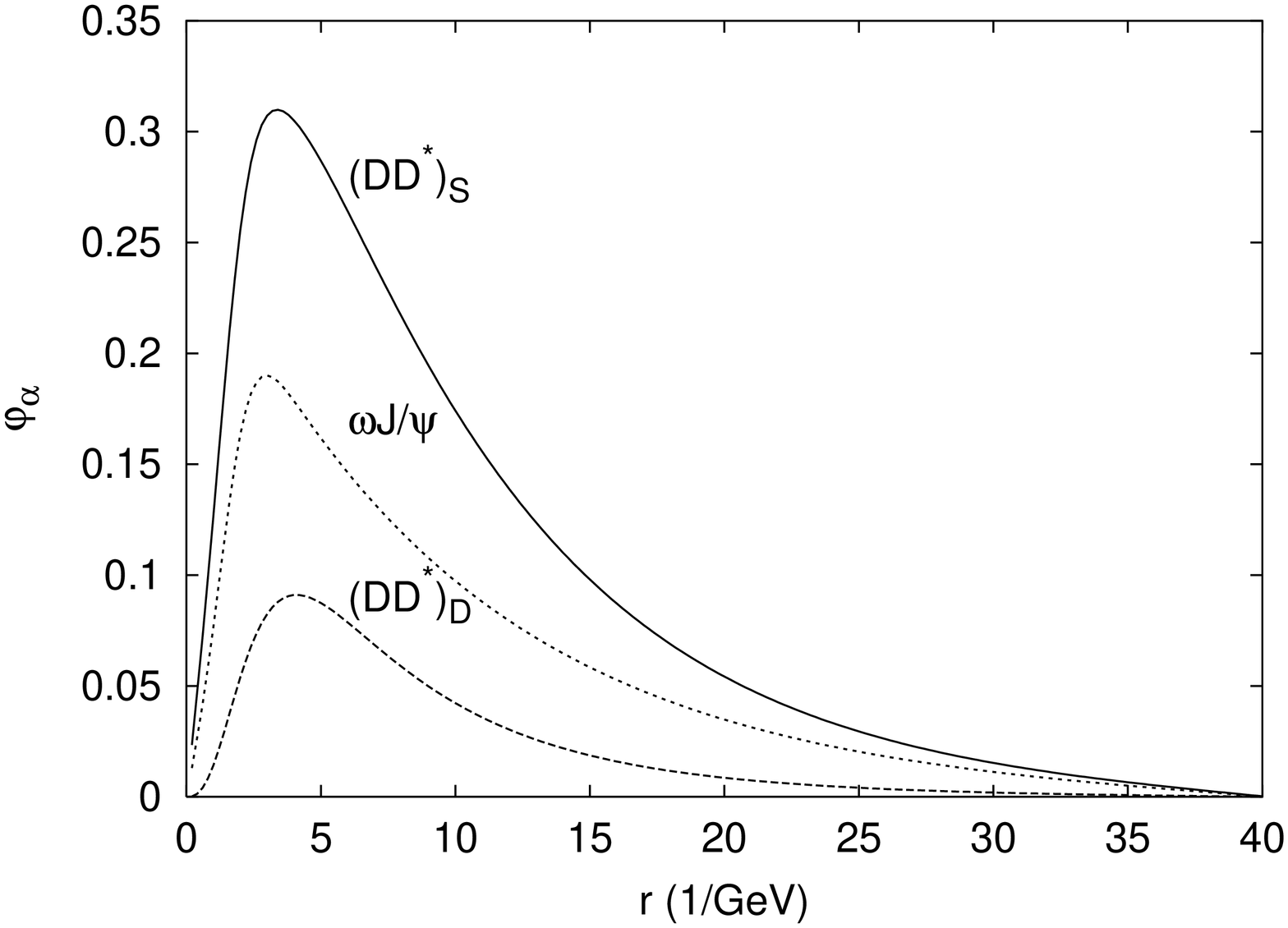}
\includegraphics[angle=270,width=8cm]{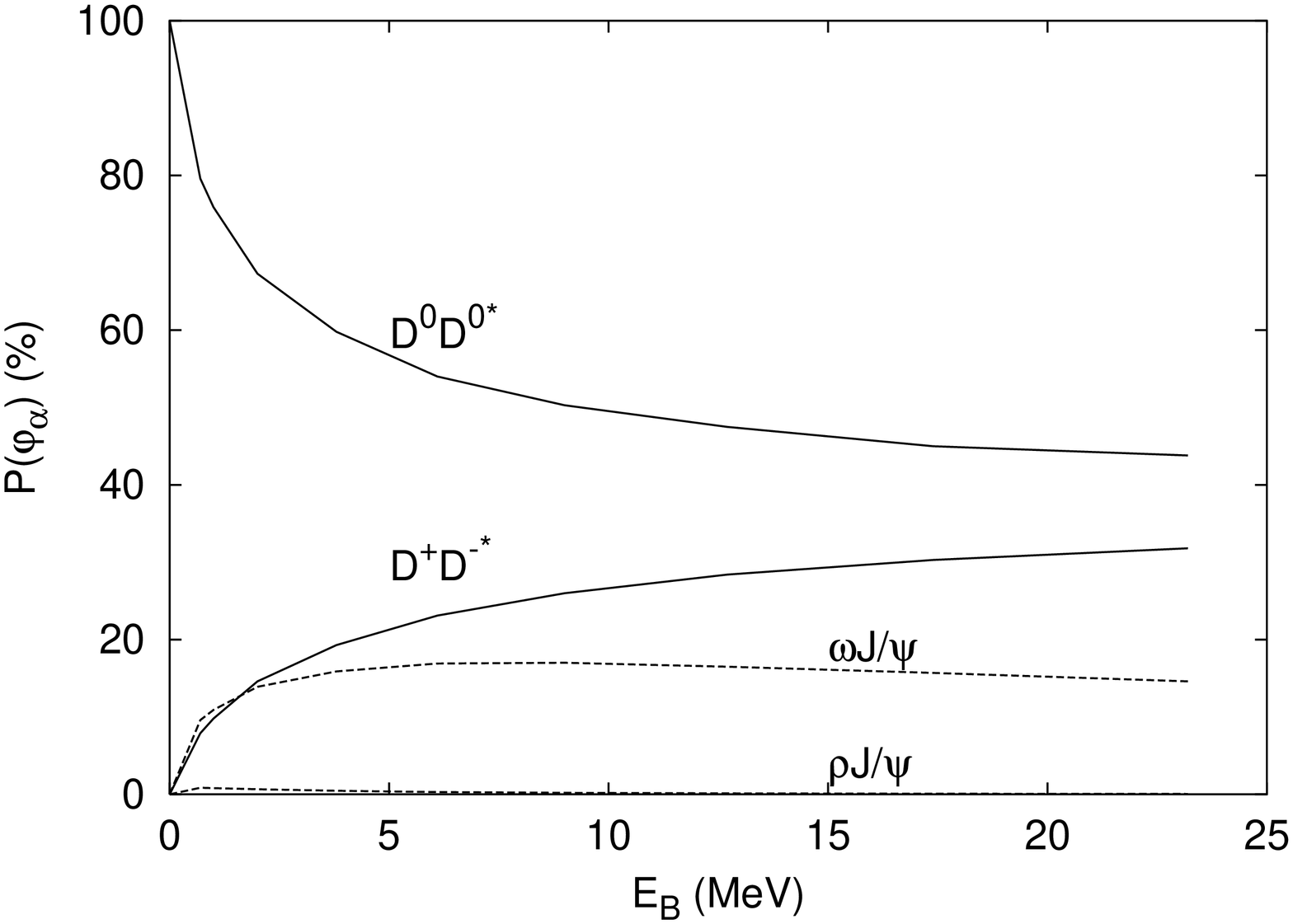}
\caption{ \label{wf} (left) Three Channel Isoscalar Wavefunction Components. (right) Component Strength vs. Binding Energy.}
\end{figure}

The proceeding considerations indicate that the isoscalar  positive charge conjugation
sector is the most likely to bind in the $D\bar D^*$ system.  Furthermore the small
branching fraction of the $\hat\chi_{c1}$ to $\pi\pi J/\psi$ implies a small isovector $\rho J/\psi$ 
component in the $\hat \chi_{c1}$ wavefunction. Thus a good initial study is provided by the coupled
channel 
$1/\sqrt{2}(D \bar D^* + \bar D D^*)^0_S$, $1/\sqrt{2}(D \bar D^* + \bar D D^*)^0_D$, 
and $\omega J/\psi$ system. Utilizing the potentials of Eqs. \ref{Vq} and \ref{Vpi} 
and meson masses of $D = 1.869$ GeV, $D^* = 2.01$ GeV, $\omega = 0.78$ GeV, and 
$J/\psi = 3.1$ GeV yields a single bound state of mass 3.872 GeV without adjusting 
any parameters, in remarkable agreement with the mass of the $X$.  The $\hat\chi_{c1}$ wavefunction
is plotted in Fig. \ref{wf} (left panel); one sees typical deuteron-like wavefunctions with 
strong D-wave and $\omega J/\psi$ components. All three components are required to 
achieve binding for this state.

Although encouraging, this result must not be taken too seriously because the binding energy
is comparable to the difference in energies of the various relevant charge channels.
Thus isospin breaking effects are expected to be important and must be incorporated
in the formalism. This is achieved by including isovector channels and allowing for
differing thresholds. Restricting attention to nearby vector meson - $J/\psi$ states
and neglecting the coupling to charmonium states\footnote{Coupling of the $\hat\chi_{c1}$ 
to charmonium 
states is negligible because the wavefunction squared at the origin scales as 
$\sqrt{\mu_{D\bar D^*} E_B}$
thereby suppressing quark annihilation transitions. Annihilation is further suppressed
because 
the $\hat\chi_{c1}$ should be dominated by S-wave quark pairs and possible 
charmonium states are all orbital excitations.}
yields six possible channels:

\begin{list}{$\bullet\ $}{}
\item{$\rho J/\psi$ at 3.8679 GeV}
\item{${1\over \sqrt{2}}(D^0 \bar D^{0*} + \bar D^0 D^{0*})_S$ at 3.8712 GeV}
\item{${1\over \sqrt{2}}(D^0 \bar D^{0*} + \bar D^0 D^{0*})_D$ at 3.8712 GeV}
\item{${1\over\sqrt{2}}(D^+D^{-*} + D^- D^{+*})_S$ at 3.8793 GeV}
\item{${1\over\sqrt{2}}(D^+D^{-*} + D^- D^{+*})_D$ at 3.8793 GeV}
\item{$\omega J/\psi$ at 3.8795 GeV}.
\end{list}

One sees an immediate problem: the threshold for $\rho J/\psi$ is too low 
to allow a resonance at $3.872\pm 1$ GeV. However, the mass of the $\rho$ is rather
poorly defined  due to its large width and some leeway in fixing
threshold for this channel is permissible.  I therefore adopt the simple 
prescription of setting the $\rho$ mass equal to that of the $\omega$ at $0.7826$ GeV.
Varying this prescription made negligible changes to the following results.
The quark level coupling of $(D\bar D^*)_D$ states to S-wave $\rho$- or $\omega$-$J/\psi$
states is small and is neglected. The final step is to form effective interactions 
from appropriate combinations of isospin basis interactions. For example the 
$D^0\bar D^{0*} - D^0\bar D^{0*}$ interaction is given by
Eq. \ref{Vpi} with $\gamma = 1$. 
The resulting numerical six channel problem must be studied with some care because
the binding energies are small relative to the natural scales of the system.  

It is of interest to study the properties of possible bound states as a function of
their binding energy. This has been achieved by allowing the regulator scale to 
vary between 1.2 and 2.3 GeV. Binding is seen to occur for $\Lambda$ larger
than approximately 1.45 GeV.
Wavefunction coefficients (defined as $\int |\varphi_\alpha|^2$ where $\alpha$ 
is a channel index) 
are shown as a function of binding energy in Fig. \ref{wf} (right panel). It is clear
that the $D^0\bar D^{0*}$ component dominates the wavefunction, especially near 
threshold.  However, the $D^+D^{-*}$ component rises rapidly in strength  with 
isospin symmetry being recovered at surprisingly small binding energies (on the order of
30 MeV). Alternatively, the $\omega J/\psi$ component peaks at roughly 17\% at $E_B \approx
 9$ MeV.
The contribution of the $\rho J/\psi$ 
wavefunction  remains small, peaking at less than 1\% very close to threshold.

%\begin{figure}[h]
%\includegraphics[angle=270,width=12cm]{r.ps}
%\caption{ \label{r} RMS radius  vs. Binding Energy.}
%\end{figure}

It is possible to estimate decay rates in a simple fashion once an explicit wavefunction
is known. This is because the small binding energy of the $\hat\chi_{c1}$ implies that its
constituent particles are nearly on-shell and therefore $\hat \chi_{c1}$ decay amplitudes
are well approximated by constituent decay amplitudes. Thus, for example, the
$\pi^+\pi^-J/\psi$ decay mode arises predominantly from the $\rho J/\psi$ wavefunction
component. The channel strengths of Fig. \ref{wf} therefore allow simple estimates
of a variety of branching fractions based on the widths of the $D^*$, $\omega$, and
$\rho$ (decays of the $D$ and $J/\psi$ mesons are neglected here but can be computed
easily). The results for a variety of modes are presented in Table I as a 
function of the binding energy. 

The broadest particle in the $\hat\chi_{c1}$ system is the $\rho$ with a width of 150 MeV and it 
is the $\rho J/\psi$ component which has the largest branching fraction, even though 
it is strongly suppressed in the wavefunction.
The next strongest mode is provided by the
$\pi^+\pi^-\pi^0$ decay of the $\omega$ which is enhanced relative to other modes due
to strong mixing with $\omega J/\psi$. 
Unfortunately, only a very rough upper limit on the total width of the $D^{0*}$ exists\cite{PDG}
so estimates of the $D^0\bar D^0 \pi^0$ and $D^0\bar D^0 \gamma$ decay widths are essentially
useless. The figures in the table have been obtained by assuming that 
$\Gamma(D^{0*}\to D^0\gamma) \approx 25$ keV and 
$\Gamma(D^{0*} \to D^0\pi^0) \approx 43$ keV; both of these estimates are
anchored in $D^{\pm*}$ decays and should be reliable. Notice that the $D^\pm \pi^\mp$ mode 
is closed.
All other possible decay modes of the $\hat\chi_{c1}$ are relatively small, although the
$\pi^0\gamma J/\psi$ mode may be of interest if it is detectable.

\begin{table}[h]
\caption{Some Decay Modes of the $\hat\chi_{c1}(3872)$ (keV).}
\begin{tabular}{cccccccccc}
\hline
$B_E$ (MeV)$^{\phantom{X}^{\phantom{X}}}$ & $D^0\bar D^0 \pi^0 \ \ $ & $D^0\bar D^0\gamma \ \ $ & $D^+D^-\pi^0 \ \ $ & $(D^+\bar D^0\pi^-$+c.c)$/\sqrt{2}$ & $D^+D^-\gamma \ \ $ & $\pi^+\pi^-J/\psi \ \ $ & $\pi^+\pi^-\gamma J/\psi \ $ & $\pi^+\pi^-\pi^0J/\psi \ \ $ & $\pi^0\gamma J/\psi \ \ $  \\
\hline
0.7  & 67 & 38 & 5.1  &  4.7  &  0.2  & 1290  & 12.9 &  720  &  70 \\
1.0  & 66 & 36 & 6.4  &  5.8  &  0.3  & 1215  & 12.1 &  820 &  80 \\
2.0  & 57 & 32 & 9.5  &  8.6  &  0.4  & 975  & 9.8 &   1040 &  100 \\
3.8  & 52 & 28 & 12.5  & 11.4  &  0.6  & 690  & 6.9  &  1190 &  115 \\
6.1  & 46 & 26 & 15.0  &  13.6 &  0.7  & 450   & 4.5 &  1270 &  120 \\
9.0  & 43 & 24  & 16.9 &  15.3 &  0.8  & 285   & 2.9  &  1280 & 125 \\
12.7 & 38 & 22  & 18.5 &  16.7 &  0.9  & 180   & 1.8 & 1240 &  120 \\ 
\hline
\end{tabular}
\end{table}

\section{Conclusions}

I have argued that the $X(3872)$ is a $J^{PC}=1^{++}$ $D\bar D^*$ hadronic resonance
with important admixtures of $\rho J/\psi$ and $\omega J/\psi$ states, dubbed the $\hat\chi_{c1}$.
 This assertion
is supported by detailed computations in a microscopic model which incorporates
pion and quark exchange interactions. The model has been heavily tested on nuclear
physics and meson-meson scattering data and can be regarded as reasonably reliable.
The $1^{++}$ $\hat\chi_{c1}$ is the only $D\bar D^*$ state which binds; no other $J^{PC}$
or charge modes exist in this model. Furthermore, no $D\bar D$ molecules are expected.
It is likely, however, that a  rich 
$D^*\bar D^*$, $B\bar B^*$ and $B^*\bar B^*$ spectrum exists.
Thus the discovery of the $X(3872)$ may be the entree into a new regime of hadronic 
physics which will offer important insight into the workings of strong QCD and 
should help clarify many open issues in light quark spectroscopy. Indeed, the experimental
and theoretical analysis of heavy molecules is  simplified because of their 
hidden flavour components.

It is clear that further experimental studies of the $X(3872)$ are of great importance.
For example, determining its spin and parity are of immediate concern. The fact that the
$X$ is polarized in $B\to KX$ will help greatly in this.  Furthermore, detecting a 
$\pi^0\pi^0J/\psi$ decay mode would immediately eliminate the $\hat\chi_{c1}$ 
interpretation of the $X$. 

It is also
important to gather enough events to reconstruct the invariant mass of various 
subsystems such as $\pi^+\pi^-$ in $\pi^+\pi^-J/\psi$ (which should peak at the 
$\rho$ mass). Perhaps a more interesting test would be the invariant mass distribution
of the $\pi^+\pi^-\pi^0$ subsystem in the $\pi^+\pi^-\pi^0J/\psi$ decay mode, which 
should have all of its events near the edge of phase space due to the narrow width
of the virtual $\omega$. It is therefore encouraging that the $3\pi J/\psi$ decay mode is 
roughly 1/2 the strength of the $2\pi J/\psi$ mode. Although some events will be lost 
due to the decreased efficiency in detecting neutral pions, this deficit should be
made up by the new data being collected at Belle and BaBar.

Lastly, determining branching fractions, especially those arising from 
different wavefunction
components such as $D^0\bar D^0 \pi^0$, $\pi^+\pi^- J/\psi$, and $\pi^+\pi^-\pi^0 J/\psi$,
would help greatly in pinning down the internal structure of the $X$ and provide an
intriguing glimpse into a new realm of hadronic physics.

\begin{acknowledgments}
I thank Jim Mueller for a fruitful conversation.
This work was supported by the DOE under contracts DE-FG02-00ER41135 and 
DE-AC05-84ER40150.
\end{acknowledgments}

\end{document}